\documentclass[traditabstract,longauth]{aa}

\usepackage{pbox}
\usepackage{graphicx}
\usepackage{txfonts}
\usepackage{natbib}
\usepackage[normalem]{ulem}

\usepackage{hyperref}	
\hypersetup{colorlinks=true,linkcolor=blue,citecolor=blue,filecolor=blue,urlcolor=blue}
\usepackage{amssymb}
\usepackage{microtype}
\usepackage{floatrow}
\DeclareFloatSeparators{myfill}{\hskip.001\textwidth plus1fill}
\usepackage[all]{nowidow}

\usepackage{afterpage}

\usepackage{needspace}

\begin{document}

   \title{LOFAR MSSS: Discovery of a 2.56 Mpc giant radio galaxy\\associated with a disturbed galaxy group}
   \titlerunning{Discovery of a giant radio galaxy in MSSS}


   \author{A. O. Clarke \inst{1},
G. Heald\inst{2,3,4},
T. Jarrett\inst{5},
J. D. Bray\inst{1},
M. J. Hardcastle\inst{6},
T. M. Cantwell\inst{1},
A. M. M. Scaife\inst{1},
M. Brienza\inst{3,4},
A. Bonafede\inst{7},
R. P. Breton\inst{1},
J. W. Broderick\inst{3},
D. Carbone\inst{8},
J. H. Croston\inst{9},
J. S. Farnes\inst{10},
J. J. Harwood\inst{3},
V. Heesen\inst{11},
A. Horneffer\inst{12},
A. J. van der Horst\inst{13},
M. Iacobelli\inst{14},
W. Jurusik\inst{15},
G. Kokotanekov\inst{16},
J. P. McKean,\inst{3}
L. K. Morabito\inst{17},
D. D. Mulcahy\inst{1},
B.S. Nikiel-Wroczy{\~n}ski\inst{15},
E. Orr\'u\inst{3},
R. Paladino\inst{18},
M. Pandey-Pommier\inst{19},
M. Pietka\inst{17},
R. Pizzo\inst{3},
L. Pratley\inst{20,21},
C. J. Riseley\inst{1,2},
H. J. A. Rottgering\inst{22},
A. Rowlinson\inst{8},
J. Sabater\inst{23},
K. Sendlinger\inst{24},
A. Shulevski\inst{3},
S. S. Sridhar\inst{3,4},
A. J. Stewart\inst{17},
C. Tasse\inst{25},
S. van Velzen\inst{26},
R. J. van Weeren\inst{27},
M. W. Wise\inst{3}}
   \authorrunning{Clarke et al.}
   \offprints{A. O. Clarke, \email{alex.clarke-3@manchester.ac.uk}}

   \institute{University of Manchester, Jodrell Bank Centre for Astrophysics, Manchester, UK, M139PL \and 
   CSIRO Astronomy and Space Science, 26 Dick Perry Avenue, Kensington WA 6151, Australia \and 
   ASTRON, the Netherlands Institute for Radio Astronomy, Postbus 2, 7990 AA, Dwingeloo, The Netherlands \and 
   Kapteyn Astronomical Institute, University of Groningen, PO Box 800, 9700 AV, Groningen, The Netherlands \and 
   Astronomy Dept, University of Cape Town, Rondebosch 7701, RSA \and 
   Centre for Astrophysics Research, School of Physics, Astronomy and Mathematics, University of Hertfordshire, College Lane, AL10 9AB \and 
   Hamburger Sternwarte - Hamburg University, Gojenbergsweg 112, 21029 Hamburg, Germany \and 
   Anton Pannekoek Institute for Astronomy, University of Amsterdam, Science Park 904, 1098XH, Amsterdam, The Netherlands \and 
   School of Physics and Astronomy, University of Southampton, Southampton SO17 1BJ \and 
   Department of Astrophysics/IMAPP, Radboud University, PO Box 9010, NL-6500 GL Nijmegen, the Netherlands. \and 
   Hamburger Sternwarte, Hamburg University, Gojenbergsweg 112, 21029 Hamburg, Germany \and 
   Max-Planck-Institut fur Radioastronomie, Auf dem Hugel 69, 53121 Bonn, Germany \and 
   Department of Physics, The George Washington University, 725 21st Street NW, Washington, DC 20052, USA \and 
   ASTRON, the Netherlands Institute for Radio Astronomy, Postbus 2, 7990 AA, Dwingeloo, The Netherlands \and 
   Astronomical Observatory, Jagiellonian University, ul. Orla 171, PL 30-244 Krakow, Poland \and 
   Anton Pannekoek Institute for Astronomy, University of Amsterdam, Science Park 904, 1098XH, Amsterdam, The Netherlands \and 
   Astrophysics, Department of Physics, University of Oxford, Keble Road, Oxford OX1 3RH, UK \and 
   INAF- Istituto di Radioastronomia, via P. Gobetti, 101, 40129 Bologna, Italy \and 
   Univ Lyon, Univ Lyon1, Ens de Lyon, CNRS, Centre de Recherche Astrophysique de Lyon UMR5574, 9 av Charles André, F- 69230, Saint-Genis-Laval, France \and 
   Mullard Space Science Laboratory (MSSL), University College London (UCL), Holmbury St Mary, Surrey RH5 6NT, UK \and 
   School of Chemical and Physical Sciences, Victoria University of Wellington, PO Box 600, Wellington 6140, New Zealand \and 
   Leiden Observatory, Leiden University, P.O. Box 9513, 2300 RA Leiden, The Netherlands \and 
   Institute for Astronomy (IfA), University of Edinburgh, Royal Observatory, Blackford Hill, EH9 3HJ Edinburgh, U.K. \and 
   Astronomisches Institut der Universitat Bochum, Universitatsstr. 150, 44801, Bochum, Germany \and 
   GEPI, Observatoire de Paris, CNRS, Universite Paris Diderot, 5 place Jules Janssen, 92190, Meudon, France \and 
   Department of Physics and Astronomy, The Johns Hopkins University, Baltimore, MD 21218, USA \and 
   Harvard-Smithsonian Center for Astrophysics, 60 Garden Street, Cambridge, MA 02138, USA 
   }

  \abstract{
  We report on the discovery in the LOFAR Multifrequency Snapshot Sky Survey (MSSS) of a giant radio galaxy (GRG) with a projected size of $2.56 \pm 0.07$ Mpc projected on the sky. It is associated with the galaxy triplet UGC\,9555, within which one is identified as a broad-line galaxy in the Sloan Digital Sky Survey (SDSS) at a redshift of $0.05453 \pm 1 \times 10^{-5} $, and with a velocity dispersion of $215.86 \pm 6.34$ km/s. From archival radio observations we see that this galaxy hosts a compact flat-spectrum radio source, and we conclude that it is the active galactic nucleus (AGN) responsible for generating the radio lobes. The radio luminosity distribution of the jets, and the broad-line classification of the host AGN, indicate this GRG is orientated well out of the plane of the sky, making its physical size one of the largest known for any GRG. Analysis of the infrared data suggests that the host is a lenticular type galaxy with a large stellar mass ($\log~\mathrm{M}/\mathrm{M}_\odot = 11.56 \pm 0.12$), and a moderate star formation rate ($1.2 \pm 0.3~\mathrm{M}_\odot/\mathrm{year}$). Spatially smoothing the SDSS images shows the system around UGC\,9555 to be significantly disturbed, with a prominent extension to the south-east. Overall, the evidence suggests this host galaxy has undergone one or more recent moderate merger events and is also experiencing tidal interactions with surrounding galaxies, which have caused the star formation and provided the supply of gas to trigger and fuel the Mpc-scale radio lobes.}

  \keywords{Surveys --- Radio continuum: galaxies --- Infrared: galaxies --- quasars: supermassive black holes --- Galaxies: interactions --- Galaxies: groups: general --- Galaxies: jets }

\maketitle

\section{Introduction}

Radio galaxies host jets of relativistic particles that originate from supermassive black holes (SMBH), usually at the centres of large elliptical galaxies. These SMBH accrete material from within the nucleus of the galaxy, and a system where accretion is taking place at a significant rate is defined as an active galactic nucleus (AGN). However, only a fraction of such AGN produce these jets, which release huge amounts of energy along a preferential axis, extending hundreds of kpc out into the intergalactic medium (IGM). In the case of giant radio galaxies (GRG), it is commonly defined that these lobes have a total projected linear size of more than 1 Mpc.

GRG are rare objects, with favourable arguments as to why they are so large suggesting that they have grown in low-density environments \citep[e.g.,][]{kaiser_etal_1997}. As well as studying individual objects, finding samples of GRG is particularly important for classifying the properties of their formation and the evolution of radio sources. The statistics of giant sources can be used to test dynamical models of radio-galaxy evolution and to probe the distribution of gas in the intergalactic medium, including potentially acting as a probe of the warm-hot intergalactic medium \citep[e.g.][]{sub2008, peng2015}.

Samples of GRG are primarily collected by searching surveys performed with radio telescopes. One example is the NRAO VLA Sky Survey \citep[NVSS;][]{condon_etal_1998}, a 1.4~GHz continuum survey covering the entire sky north of \mbox{$-40$} degrees declination. From this survey, \cite{Solovyov_etal_2011} compiled a sample of 61 new GRG candidates. Another example is the Westerbork Northern Sky Survey \citep[WENNS;][]{WENNS} at 325 MHz above +28 degrees declination, where \cite{Schoenmakers_2001} found 47 GRG candidates. However, the size of any sample of extended objects is dependent on the ability to image the large scale, low surface brightness emission of sources such as GRG, which is highly dependent on the \emph{uv}-coverage and observing frequency of the telescope.

The Low Frequency Array \citep[LOFAR;][]{LOFAR} is a well suited instrument for detecting GRG and exploring their morphology. It can produce arcsecond resolution maps, has a wide field of view, and has excellent \emph{uv}-coverage which makes it sensitive to the large angular scales of nearby GRG. Moreover, due to the nature of synchrotron emission, these objects are brighter at the low frequencies (150~MHz) where LOFAR operates. This is especially important when considering the evolutionary models of giant radio sources discussed in \cite{kaiser_etal_1997}, as they expect GRG to be older objects that have a lower luminosity than smaller radio galaxies. \cite{hardcastle2016} discuss the discovery of giant sources with LOFAR in their survey of the H-ATLAS field.

This paper is organised as follows. In Section \ref{section:data} we present the discovery of a new GRG in the quality control imaging produced as part of the LOFAR Multifrequency Snapshot Sky Survey \citep[MSSS;][]{heald_etal_2015}, along with ancillary radio, optical and infrared data collected from various sources in the literature, and summarise key properties of the host galaxy in table \ref{table:u9555}. In Section \ref{section:discussion}, we discuss the physical nature of this GRG, along with an analysis of the host system. We summarise the main conclusions in Section \ref{section:conclusion}. Throughout this paper we adopt the spectral index convention that $S\propto\nu^\alpha$, where S is the flux density, $\nu$ is the frequency, and $\alpha$ is the spectral index. We assume a flat universe with \mbox{${\Omega_\mathrm{m}} = 0.286$}, \mbox{${\Omega_\Lambda} = 0.714$}, and \mbox{${\textrm{H}}_0 = 67.8\,\textrm{km} \,\textrm{s}^{-1} \textrm{Mpc}^{-1}$} \citep{plank2015}, unless explicitly stated otherwise.

\section{Data and Analysis}\label{section:data}

\subsection{LOFAR MSSS}

\begin{figure*}
\includegraphics[width=\hsize]{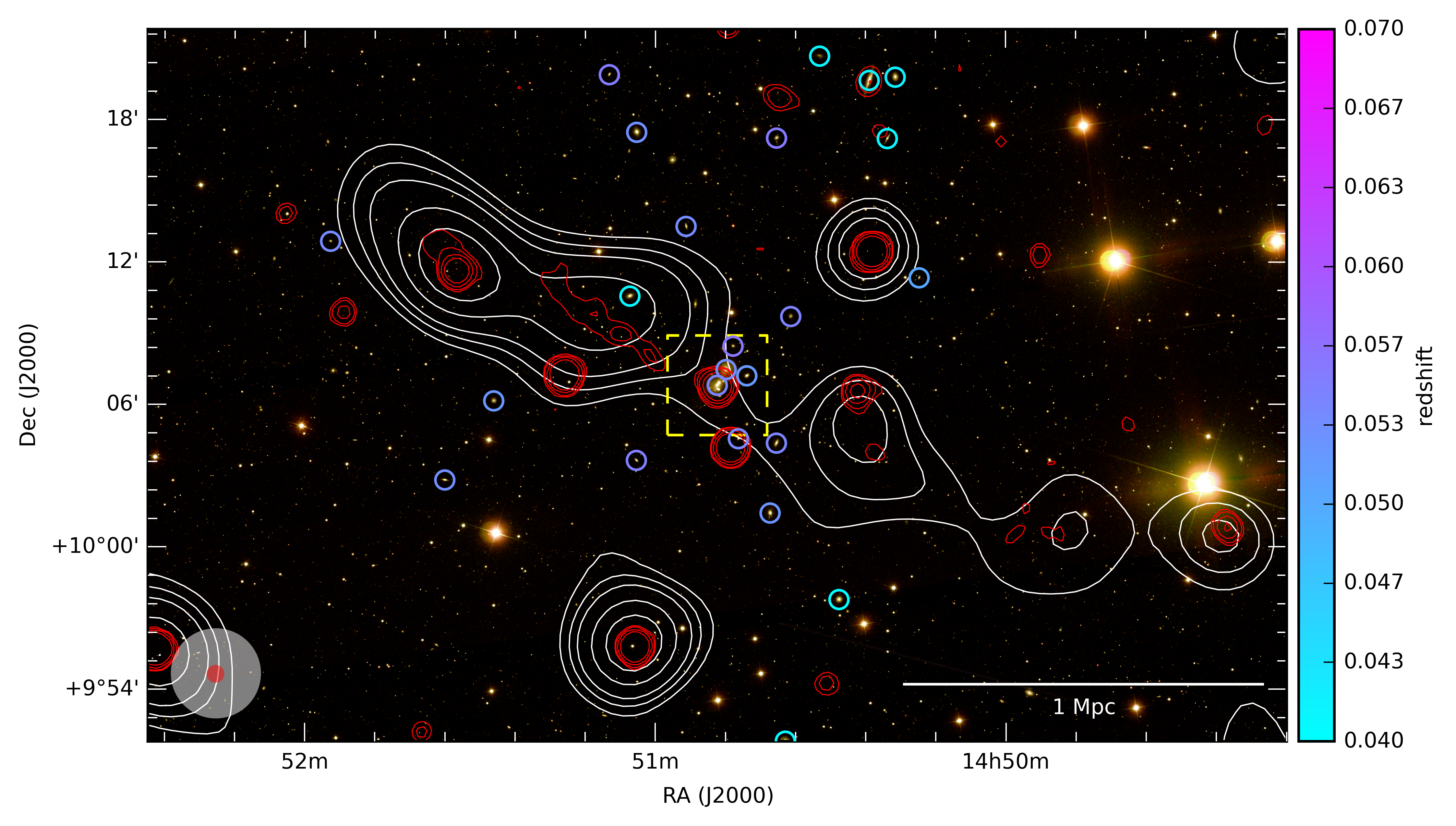}
\caption{Background SDSS image (composite from bands \textit{g}, \textit{r} and \textit{i}) overlaid with white MSSS contours of the GRG at 2, 3, 4, 6 and 8 times the RMS noise (34 mJy/beam). NVSS contours are overlaid in red at 3, 5, 10 and 20 times the RMS noise (0.55 mJy/beam) revealing a bright part of the radio jet towards the north-east. The beam sizes are shown in the lower left. Galaxies with spectroscopic detections are marked with circles coloured by their corresponding redshift, where the AGN lies at a redshift of $0.05453 \pm 1 \times 10^{-5}$. The dashed yellow box shows the area of Figure \ref{figure:sdss_smoothed_zoom}.}
\label{figure:msssnvss}
\end{figure*}

The LOFAR survey, MSSS, covers the sky north of the celestial equator, at frequencies from 119--158 MHz in 8 separate 2 MHz bands. Images are provided for each band, as well as an average of the whole bandwidth. The targeted resolution is approximately $2 \arcmin$, with a targeted root mean squared (RMS) noise of 10 mJy/beam in each band, achieved with two 7-minute scans separated by 4 hours to improve \emph{uv} coverage. Full details are given by \citet{heald_etal_2015}.

The LOFAR MSSS data for fields covering this GRG were collected on 2013-02-22 and 2013-07-28. Each of the 8 bands was imaged with an inner \emph{uv} cut of $0.1~\rm{k \lambda}$, and an outer \emph{uv} cut of $2~\rm{k\lambda}$, and combined into a $10^\circ\times10^\circ$ mosaic. For this paper, we took a weighted average of these 8 single-band mosaics to minimise the RMS noise, producing a full-band mosaic with a central frequency of 142 MHz, a resolution of $3.8\arcmin$, and RMS noise of 30 mJy/beam. Figure~\ref{figure:msssnvss} shows MSSS contours from this mosaic in the region of the GRG, along with NVSS contours overlaid on a composite Sloan Digital Sky Survey \citep[SDSS;][]{SDSS} background from bands \textit{g}, \textit{r} and \textit{i}. We measure the total angular size of the extended MSSS structure to be $39^\prime \pm 1\arcmin$. Anticipating the optical and higher frequency radio information given in Section \ref{subsection:hifreqradio} and Section \ref{subsection:SDSS}, we identify the AGN responsible for the Mpc radio emission at a redshift of $0.05453 \pm 1 \times 10^{-5}$, which puts the projected linear size of the GRG at $2.56 \pm 0.07$ Mpc. The uncertainty on this size is calculated from the uncertainty on the angular size of the object, and assumes no uncertainty on $\rm{H_0}$, and that the uncertainty on the redshift is negligible. 

The flux density calibration for MSSS requires an additional bootstrapping method in order to correct the flux scale. This is because the procedure of transferring gains from the calibrator to the target field introduces errors in the flux density calibration due to inadequate knowledge of the beam on the sky. This is also frequency dependent and so affects the spectral index of sources. In summary, the procedure to correct for this is done per field, and takes bright point sources to fit spectra to from each of the 8 band images. These sources are then cross-matched with detections in other surveys, in order to deduce correction factors for each MSSS field, for each band. \cite{hardcastle2016} use this method in their data, and describe the details further in their section 3.2.

The inner \emph{uv} cut used in imaging, and the sparse \emph{uv} coverage due to the snapshot (7 minutes) nature of the two observations, limit the imaging fidelity and particularly flux estimation for this large, extended object. The inner \emph{uv} cut leads to a central hole in coverage of the \emph{uv} plane with a diameter of $200~\lambda$, corresponding to a partial loss of sensitivity to structures with scales $\gtrsim 1/200~\rm{rad} = 17\arcmin$, roughly the size of each of the lobes of this GRG. A further complication is that at this resolution we are not able to disentangle the contribution from unassociated point sources overlapping the extended emission, as seen in NVSS. Considering these limitations of the MSSS data for such a target, we give a preliminary estimate of the integrated flux of this GRG to be $1.54 \pm 0.2~\rm{Jy}$ in this 142 MHz MSSS image, measured from the $2\sigma$ contour. At a luminosity distance of 251.3 Mpc, this gives a total luminosity at 142 MHz of $1.16 \times 10^{25}$ W/Hz.

{\def\arraystretch{1.2}\tabcolsep=3pt
\begin{table}
\caption{Properties of the host galaxy (UGC\,9555 NED03). The age and mass from SDSS is not presented with an uncertainty in their catalogue. The 8.4~GHz measurement from CLASS is not quoted with an uncertainty in their catalogue. The 1.4~GHz measurement from the FIRST is not quoted with an uncertainty in their catalogue. }
\begin{tabular}{@{}lll@{}}
Property & Value & Reference \\
\hline
RA (J2000) & 222.70582$^{\circ}$ & \cite{DR12} \\
Dec (J2000) & 10.113635$^{\circ}$ &\cite{DR12} \\
Redshift & \pbox{5cm}{0.05453 \\ $ \pm 1 \times 10^{-5} $ }& \cite{DR12} \\
Luminosity Distance (Mpc) & 251.3 & \cite{Wright_2006} \\
Age (Gyr) & 12.75 & \cite{Mar2009} \\
Velocity Dispersion (km/s) & $215.86 \pm 6.34$ & \cite{DR12} \\
Mass (SDSS) ($\log~\mathrm{M}_{\mathrm{gal}}/\mathrm{M}_\odot$) & 11.37 & \cite{Mar2009}\\
Mass (WISE) ($\log~\mathrm{M}_{\mathrm{gal}}/\mathrm{M}_\odot$) & $11.56 \pm 0.12$ & see Section \ref{sec:wise} \\
Star Formation Rate ($\mathrm{M}_\odot/\mathrm{year}$) & $1.2 \pm 0.3$ & see Section \ref{sec:wise} \\
Black Hole Mass ($\log~\mathrm{M}_{\mathrm{BH}}/\mathrm{M}_\odot$) & $8.26 \pm 0.3$ & see Section \ref{subsection:SDSS}\\
TGSS-ADR1, $\mathrm{S}_\mathrm{150 MHz}$ (mJy) & $40.1 \pm 7.6$ & \cite{intema_2016} \\ 
NVSS, $\mathrm{S}_\mathrm{1.4 GHz}$ (mJy) & $55.3 \pm 2.1$ & \cite{condon_etal_1998} \\ 
FIRST, $\mathrm{S}_\mathrm{1.4 GHz}$ (mJy) & $47.58$ & \cite{becker_etal_1995} \\
CLASS, $\mathrm{S}_\mathrm{8.4 GHz}$ (mJy) & 48.9 & \cite{myers2003}\\
\end{tabular}
\label{table:u9555}
\end{table}
}
\subsection{VLSSr}\label{subsection:hifreqradio} 

Neither the lobes of the GRG, nor the central AGN appear in The Very Large Array Low-frequency Sky Survey Redux \citep[VLSSr;][]{VLSSr} at 74~MHz. The noise in the VLSSr region is 100 mJy/beam and the jet of the GRG seen in NVSS would need a spectral index of $\alpha \leqslant -1.3$ to be visible at three times the noise. Furthermore, given that the central AGN is a flat spectrum source, we do not expect a detection in VLSSr.

\subsection{FIRST}\label{section:first}

\begin{figure*}
\includegraphics[width=0.498\hsize]{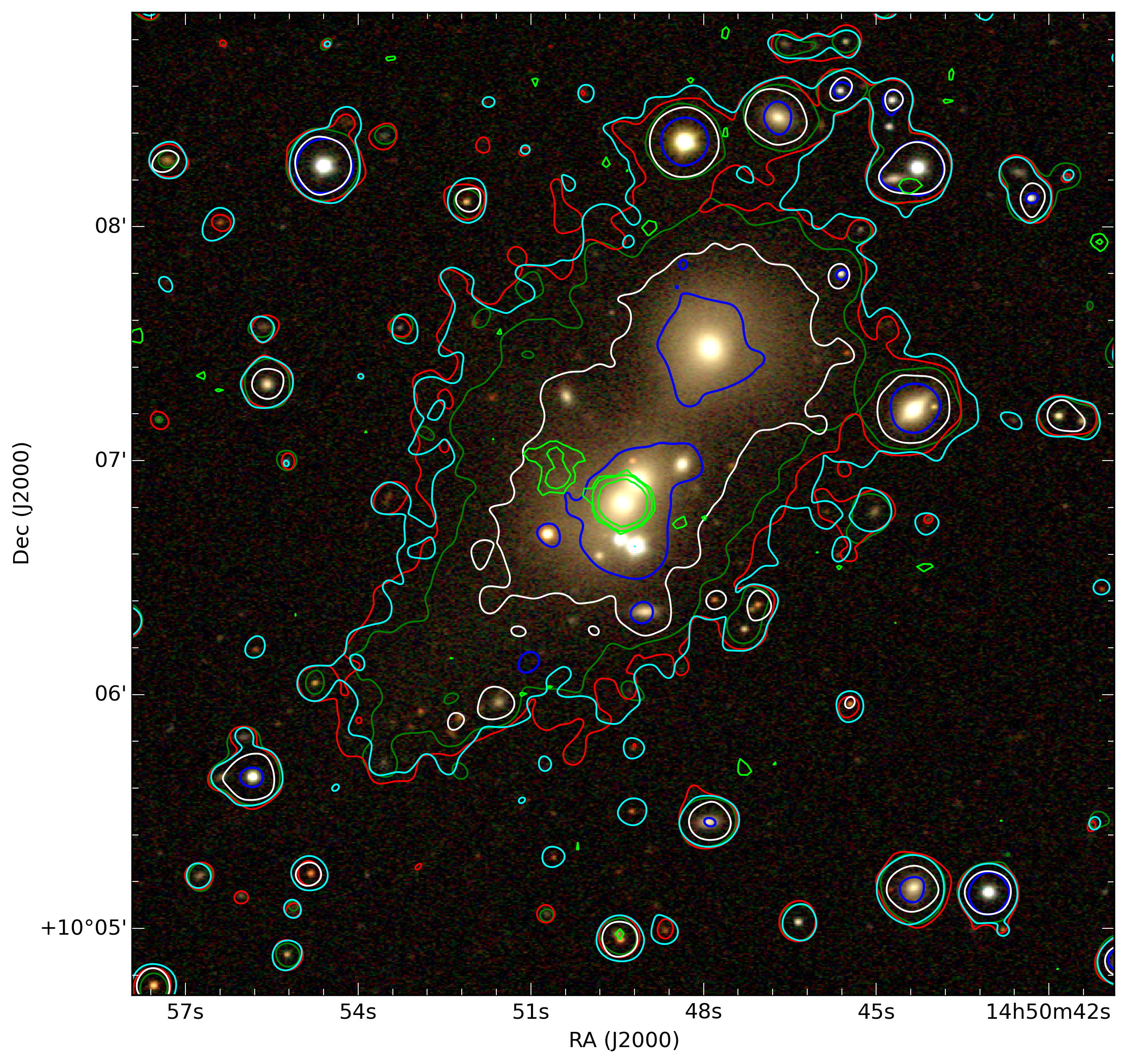}
\includegraphics[width=0.498\hsize]{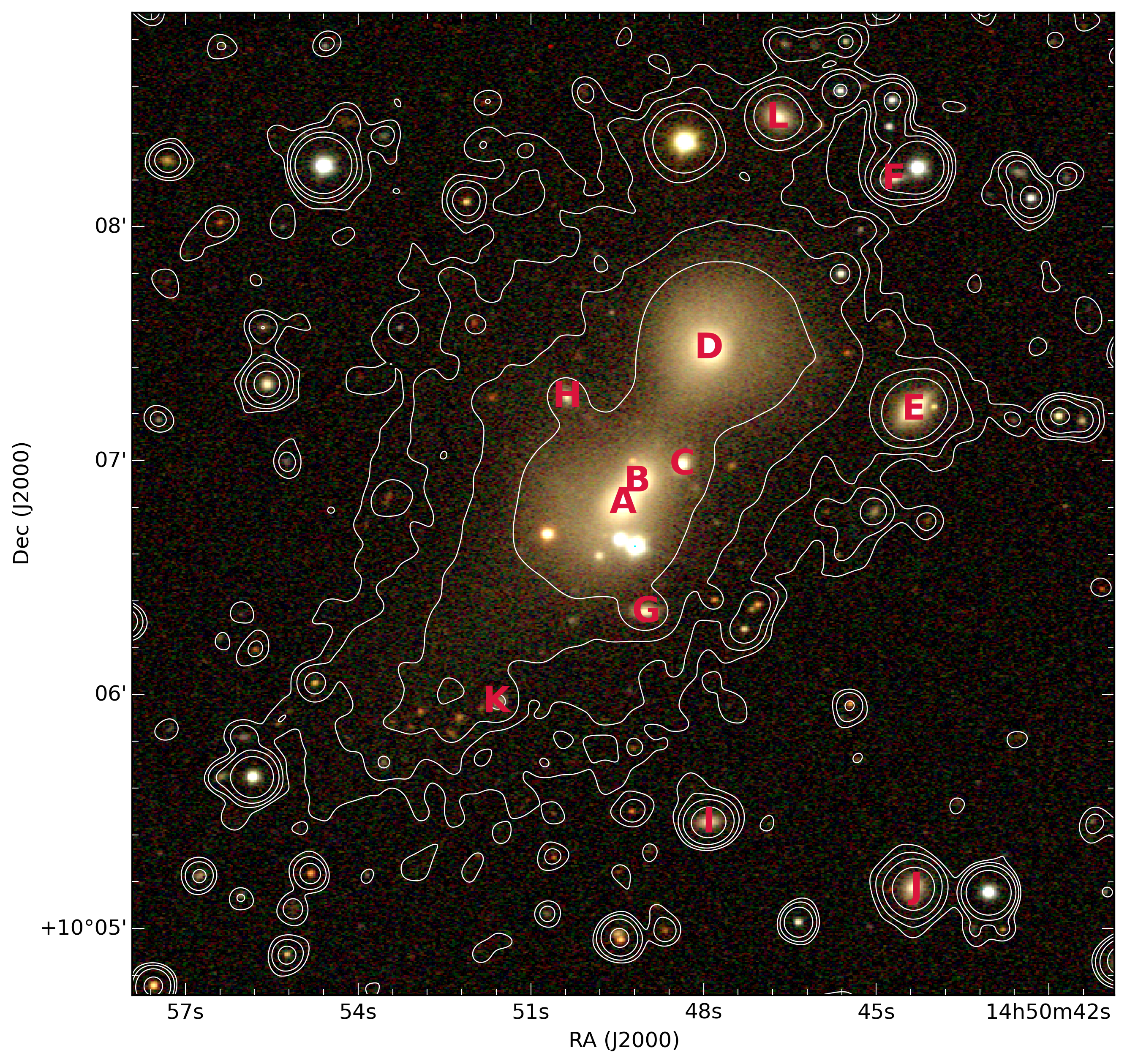}

\caption{Left: SDSS image (composite 3 colour image from bands \textit{i}, \textit{r} and \textit{g} on log stretch scale) with SDSS contours at 3 times the RMS noise after smoothing from bands \textit{z} (white), \textit{i} (cyan), \textit{r} (red), \textit{g} (green) and \textit{u} (blue). The RMS noise of each band before smoothing in units of maggies is $4.14 \times 10^{-2}$ (\textit{u}), $1.44 \times 10^{-2} $ (\textit{g}), $2.38 \times 10^{-2}$ (\textit{r}), $3.82 \times 10^{-2} $ (\textit{i}), $2.1 \times 10^{-1} $ (\textit{z}). FIRST contours are shown in lime at 3, 5 and 15 times the RMS noise. Right: SDSS image (composite 3 colour image from bands \textit{i}, \textit{r} and \textit{g} on log stretch scale) with smoothed SDSS contours from the composite image at 3, 5, 10 and 30 times the RMS noise after smoothing. Letters mark the positions of resolved galaxies. All smoothing is performed with a Gaussian kernel of standard deviation 1.72\arcsec. This area is shown by a dashed yellow rectangle in Figure \ref{figure:msssnvss}.}
\label{figure:sdss_smoothed_zoom}
\end{figure*}

In the VLA survey, Faint Images of the Radio Sky at Twenty Centimeters \citep[FIRST;][]{becker_etal_1995}, we identify a compact source at the centre of the MSSS structure, coincident with member A of the galaxy group UGC\,9555, as shown in Figure \ref{figure:sdss_smoothed_zoom}. It is compact but with an additional component to the north east, consistent with the orientation of the Mpc-scale structure identified by the MSSS data. The integrated flux density from the FIRST catalogue for this central AGN is 47.58 mJy. The component to the north-east has an integrated flux density of 2.43 mJy, also in the FIRST catalogue. Neither values are stated with uncertainties in the catalogue. The RMS noise in the map is 0.14 mJy/beam.

There is a FIRST detection of a source behind the north-east lobe, with resolved structure. We suggest that the optical counterpart for this source (J145134.03+101142.2) is a faint red galaxy (\textit{r} magnitude 20.242; \citealt{DR12}), with unknown redshift. This galaxy is $27\arcsec$ from the galaxy cluster J145135.4+101123 \citep{Wen2012} with a photometric redshift of 0.524. We suggest the radio emission in FIRST is from a background radio galaxy within this cluster at redshift 0.524, and is not related to the giant nearby structure seen in MSSS.

\subsection{NVSS}
In the NVSS survey at 1.4~GHz, the compact source at the location of the AGN detected in FIRST has an integrated flux density of $55.3 \pm 2.1$~mJy, taken from the catalogue. The RMS noise in the map is 0.55 mJy/beam. Compared to FIRST, the NVSS data show an increased integrated flux density at 1.4~GHz. Considering the larger beam size in NVSS ($45\arcsec$), the increased flux could be due to the extension of the jet out from the AGN. Whilst a significant portion of the diffuse MSSS structure is not seen in NVSS, there is notable elongated emission which extends approximately $4 \arcmin$ to the north east, but is not connected to the compact source (see Figure \ref{figure:msssnvss}). The extended NVSS emission is only half of the length of the north-east lobe in the MSSS image, coincident with the highest surface brightness part of the lobe in the MSSS data. We do not see any evidence of a jet in the south-west lobe, which could be due to relativistic Doppler boosting if the south-west jet is orientated away from us. This would suggest that the north-east lobe is orientated towards us, and the whole GRG structure is at an angle off perpendicular to our line of sight.

The FIRST source at the end of the north-east lobe (J145134.03+101142.2; discussed in Section \ref{section:first}) has some resolved structure at 3 sigma in NVSS. This could be a combination of the background source being resolved at $45\arcsec$, and structure in the GRG lobe such as a hotspot. Given the point spread function of the MSSS image, and the extent to which the lobe continues beyond this background point source, we conclude there is still diffuse emission from the lobe of the GRG extending beyond the point source seen in NVSS and FIRST.

\subsection{GMRT}
In the Tata Institute of Fundamental Research GMRT Sky Survey Alternate Data Release 1 \citep[TGSS-ADR1;][]{intema_2016} at 150 MHz and $25 \arcsec$ resolution, no emission is detected from the lobes of the GRG. We do not expect a detection of the lobes due the high resolution of the maps, and the arrays low sensitivity to large angular scales given the 15 minute integration time. A compact source coincident with the AGN is present, with no extension noticeable at this resolution. The integrated flux density of this source in their catalogue is $40.1 \pm 7.6$~mJy. This is a decrease in integrated flux density as compared to the FIRST measurement at 1.4~GHz of 47.58~mJy. Considering we expect no contamination from the lobes in the FIRST measurement, we expect this decrease in flux is due to synchrotron self-absorption in the jet, which occurs at low frequencies. The background point source at the farthermost end of the north-east jet (as discussed in the previous sections) is detected but not resolved in TGSS-ADR1.

\subsection{CLASS}\label{subsection:spectralindex}
The Cosmic Lens All-Sky Survey \citep[CLASS;][]{myers2003} observed this AGN at 8.4~GHz with the Very Large Array in A configuration, which has a beam size of roughly 0.2\arcsec. They measure a flux density of 48.9 mJy, without an uncertainty. They derive an elliptical gaussian model fit with a beam major axis of 50~mas, indicating that the source is still unresolved.

\subsection{The Sloan Digital Sky Survey}\label{subsection:SDSS}

Optical maps from SDSS on the region containing the AGN and surrounding galaxies (labeled with capital letters) are shown in Figure \ref{figure:sdss_smoothed_zoom}. Spectra from the SDSS Data Release 12 (DR12; \citealt{DR12}) identifies galaxy A (coincident with the FIRST emission) as a broad-line galaxy with a redshift of $0.05453 \pm 1 \times 10^{-5}$. It has no resolved structure, and has the appearance of an elliptical galaxy. The identification of broad lines indicates that this galaxy hosts an AGN, and is likely responsible for the Mpc-scale radio lobe emission seen in MSSS, at a luminosity distance of 251.3 Mpc from us, and viewed at intermediate inclination to the line of sight. This host galaxy has a velocity dispersion of $215.86 \pm 6.34 ~\rm{km/s}$, and the other two galaxies in this triplet (B and C) do not have spectra. There is also a bright galaxy to the north (D) that has a spectroscopic redshift of $ 0.05332 \pm 1 \times 10^{-5} $, and a velocity dispersion of $214.01 \pm 4.80 ~\rm{km/s}$.

After smoothing each SDSS band with a Gaussian kernel of standard deviation 1.72\arcsec, the region shows significant extended stellar emission, mainly stretching from the north-west to the south-east. For the smoothing, the relatively small width of the kernel was chosen to improve sensitivity to the extended structure, whilst retaining adequate resolution such that surrounding point sources were not confused with the diffuse emission. The left image in Figure \ref{figure:sdss_smoothed_zoom} shows smoothed contours from each SDSS band. The extension of the stellar distribution is clearer when showing smoothed contours from the composite \textit{i}, \textit{r} and \textit{g} image (Figure \ref{figure:sdss_smoothed_zoom}, right image). We suggest this extended stellar distribution is due to a history of mergers, and tidal interactions of the galaxy triplet (A, B, C) and galaxy D, and could also involve interactions with other surrounding galaxies (E--L).

We estimated the mass of the host SMBH using the empirical relation between the galaxies velocity dispersion and black hole mass, as described by \cite{shankar2016}. We first correct the SDSS velocity dispersion for the spectra aperture size using \cite{cappellari2006}, and use equation 7 in \citeauthor{shankar2016} to find $\log~\mathrm{M}_{\mathrm{BH}}/\mathrm{M}_\odot = 8.26 \pm 0.3$. The error is dominated by the scatter in the empirical relation (\citeauthor{shankar2016}, upper left panel of Figure 4); the uncertainty resulting from the velocity dispersion measurement is only $\pm 0.01$.

\afterpage{%

\begin{table*}
\centering
\caption{Properties of the local cluster identifications from multiscale probability mapping in \cite{MSPM2012}.}
\begin{tabular}{p{90mm}|l|l|l}
Property & MSPM02158 (UGC~9555) & MSPM01932 & MSPM01752 \\
\hline
Right Ascension (J2000) & 222.7058$^{\circ}$ & 222.6143$^{\circ}$ & 222.185$^{\circ}$ \\
Declination (J2000) & 10.1136$^{\circ}$ & 9.5111$^{\circ}$ & 11.3247$^{\circ}$ \\
Average redshift from group members & 0.05484 & 0.05052 & 0.05126 \\
Count of galaxies with \textit{r} mag $<$ 17.77 & 8 & 18 & 33 \\
Local density contrast (from 0.4 to 2 Mpc) & 16.6 & 22.6 & 7.9 \\
Local density contrast (from 1 to 2 Mpc) & 5.2 & 4.6 & 4.8 \\
Velocity dispersion (km/s) & 235 & 65 & 466 \\
Galaxy density within 0.4 Mpc (units of the background density) & 123.2 & 190.1 & 225.1
\end{tabular}
\label{table:MSPM}
\end{table*}

\begin{figure*}[h!]

\includegraphics[width=\hsize]{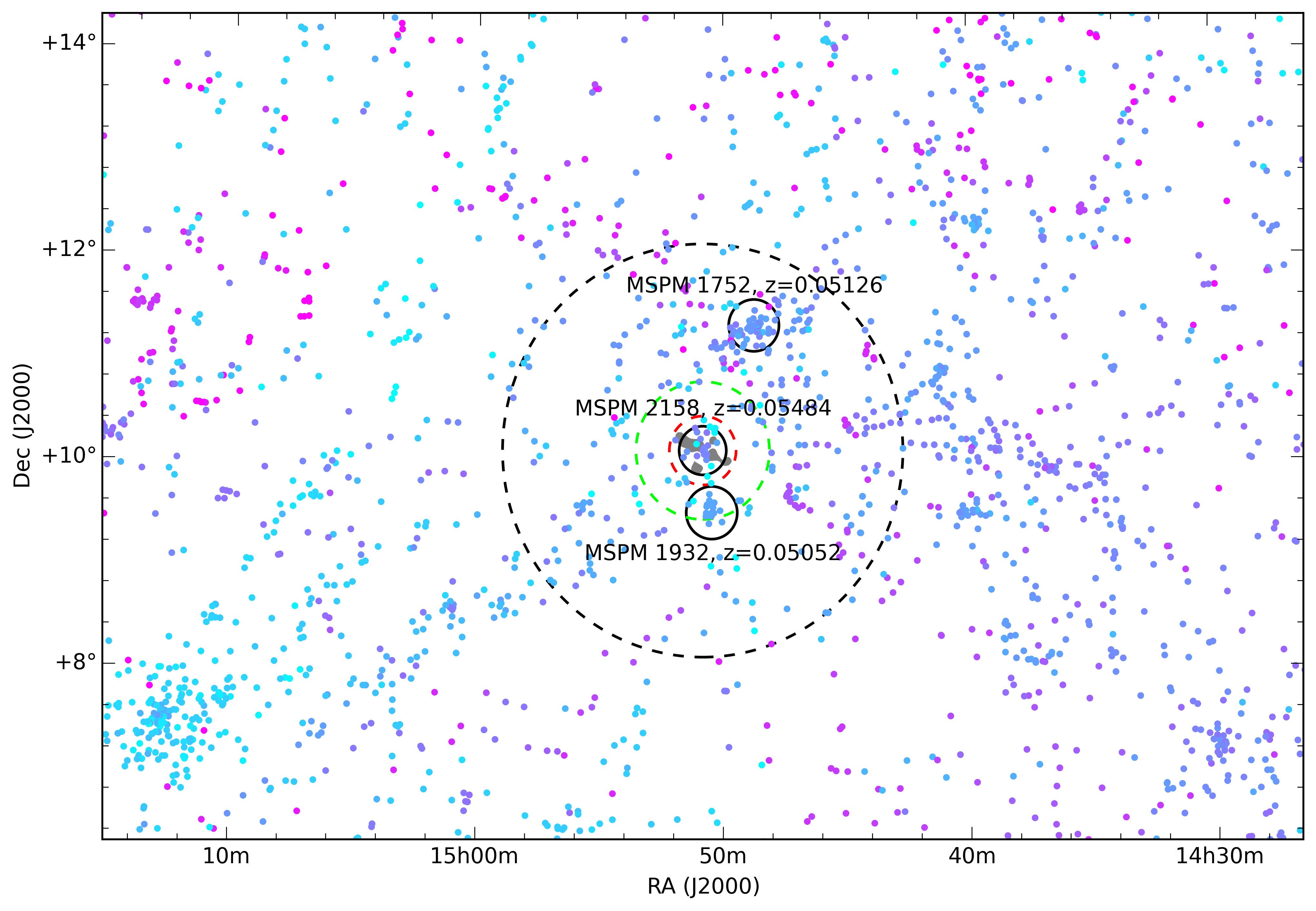}

\floatbox[{\capbeside\thisfloatsetup{capbesideposition={right,center}}}]{figure}[\hsize]
{\caption{Top: The large-scale galaxy distribution is shown, where filled circles coloured by redshift are galaxies identified with spectroscopic redshifts in SDSS DR12. Black circles of 1 Mpc radius (at their respective redshifts) are centred on the closest three galaxy groups in the region identified by multiscale probability mapping analysis. The grey-shaded area in MSPM~2158 shows the extent of the GRG. For a detailed view of the inner region around the GRG, see Figure \ref{figure:msssnvss}. Dashed circles in black, green and red are overlaid to show the areas included in the histogram to the left. Left: A histogram of the redshifts of galaxies in SDSS Data Release 12 within several radii around the host AGN, as illustrated in the top panel.}
\label{figure:redshift_hist_image}}
{\includegraphics[width=\hsize]{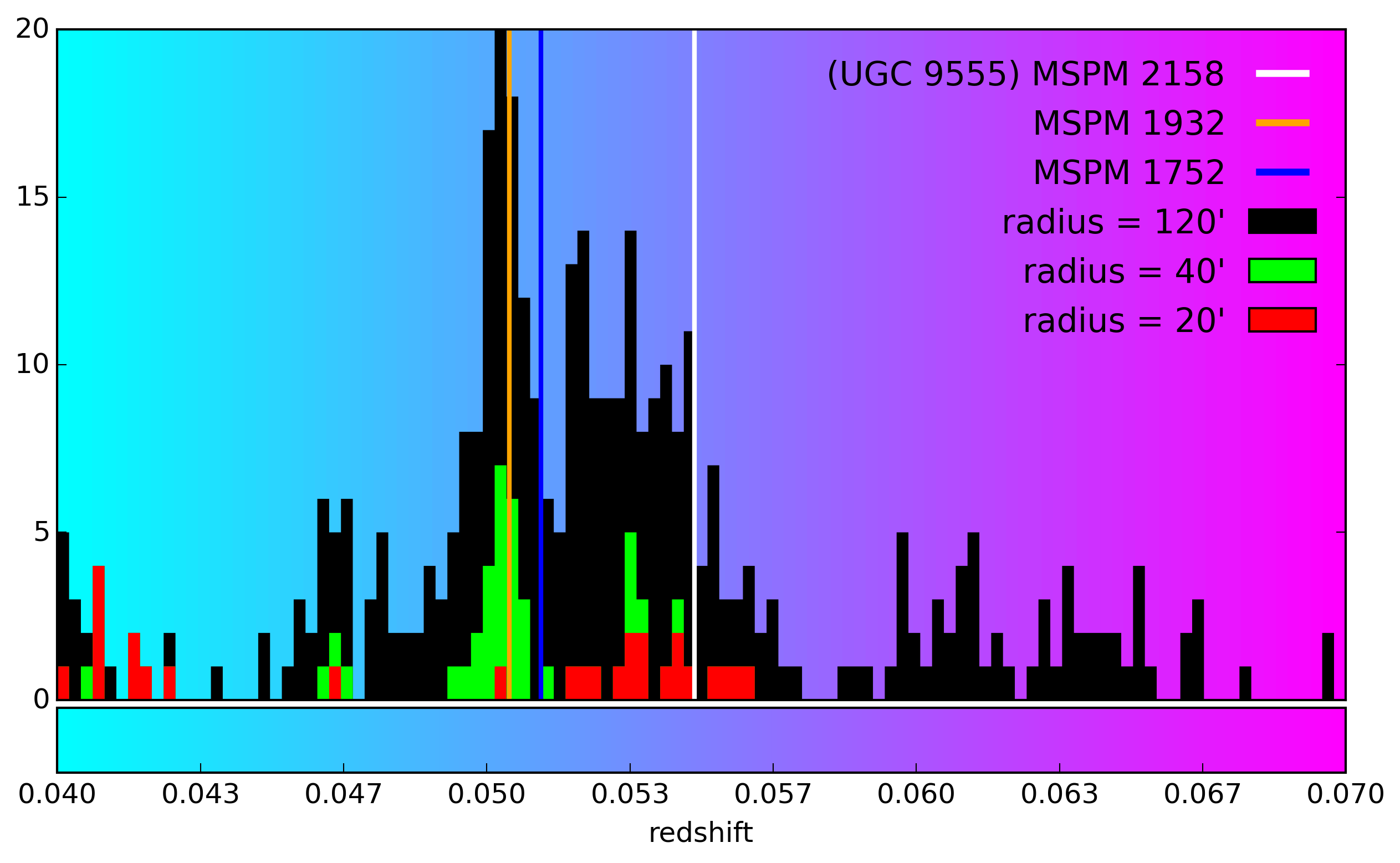}}

\end{figure*}
\clearpage
}

\subsection{The large scale environment}\label{subsubsection:largescaleenv}

We use SDSS data to constrain properties of the large-scale environment surrounding the AGN. A study done by \cite{MSPM2012} on multiscale probability mapping (MSPM) of SDSS Data Release 7 identifies UGC\,9555 as part of a larger group of galaxies they define as a cluster (MSPM 02158). They also identify two other larger galaxy clusters nearby which dominate the mass density in the region. Properties of these galaxy clusters are given in Table \ref{table:MSPM}, and Figure \ref{figure:redshift_hist_image} shows their location on the sky. The larger cluster (MSPM 01752) is 1.5 degrees (5.9 Mpc) to the north-west of UGC\,9555, and at a redshift of 0.05126 (5.9 Mpc closer to us along the line of sight), giving a total physical separation of 13.9 Mpc. The smaller cluster (MSPM 01932) is $35\arcmin$ (2.3 Mpc) to the south of UGC\,9555, at a redshift of 0.05052 (15.5 Mpc closer), with total physical separation of (15.7 Mpc). The cluster identified as hosting the AGN (MSPM 02158) has the fewest members, and has a lowest galaxy density within 0.4 Mpc than the other two clusters.

The MSPM work also searched for filaments and voids, none of which are identified in this redshift range. However, Figure \ref{figure:redshift_hist_image} suggests that the AGN lies at a junction of converging filamentary structures, with apparent under-densities in the area as well. A filamentary structure is particularly prominent going from the AGN to the south-east towards a much larger cluster of galaxies in the foreground. There are also large extended structures to the west of the AGN. 

Figure \ref{figure:redshift_hist_image} also shows a histogram of the distribution of galaxies with spectroscopic redshifts, within several radii around the AGN. Looking within a 2 degree radius of the AGN, there is a broad distribution of galaxies across a large redshift range, with the peak at the redshift of MSPM 01932. Despite the large galaxy count around the redshift of the AGN in UGC\,9555, the distribution is predominantly beyond the reach of the radio lobes (becoming significant at a radius of $40 \arcmin$ to $120 \arcmin$, equivalent to 2.6 to 7.9 Mpc at the redshift of the AGN). The region within a $20\arcmin$ radius around the AGN has 17 galaxies (with redshifts from 0.050 to 0.056 in Figure \ref{figure:redshift_hist_image}), which lie predominantly in a north-south distribution, leaving under-densities in the east-west direction where the radio lobes extend.

There are no X-ray data for this source, and we do not see a detection in the ROSAT all-sky X-ray survey. The velocity dispersion from Table \ref{table:MSPM} implies a group environment with an X-ray temperature that would be expected to be around 1.0 keV \citep{helsdon2000}. This is a reasonably typical environment for a nearby, low-power radio galaxy \citep{croston2008}. Follow-up X-ray observations would be needed to confirm this in detail, however.

\begin{figure*}[]
\includegraphics[width=0.498\hsize]{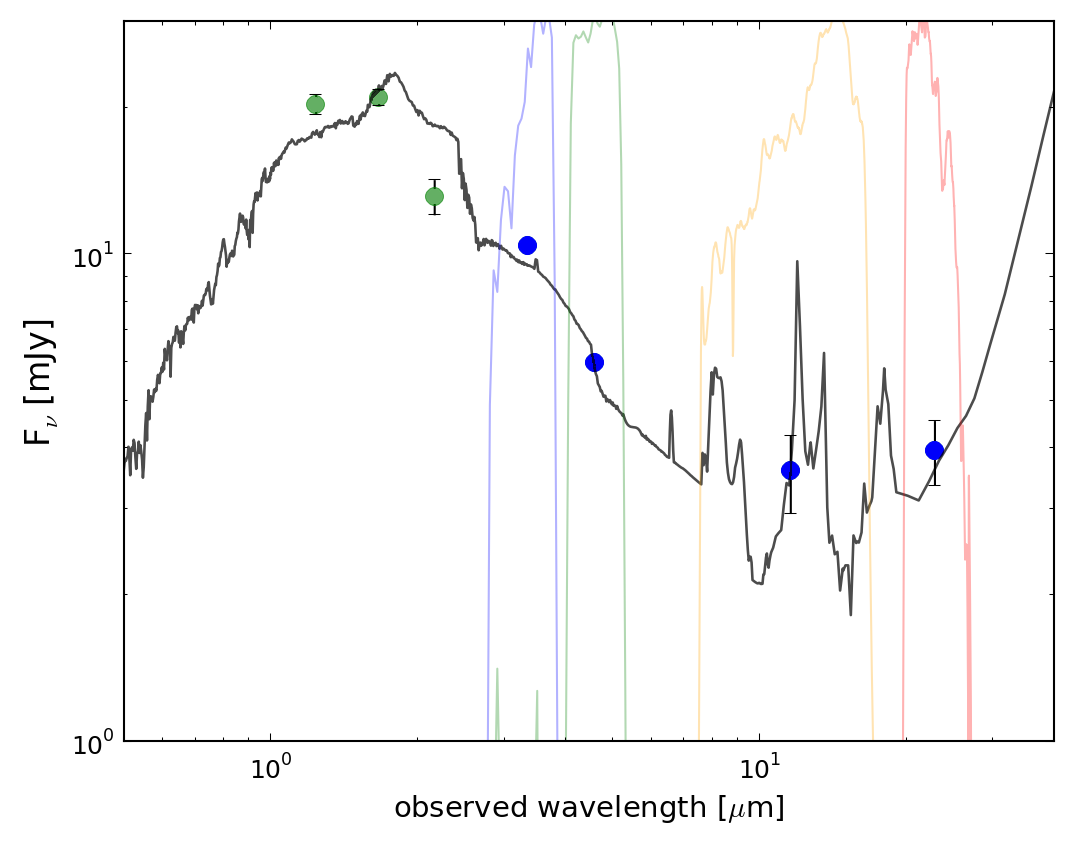}
\includegraphics[width=0.498\hsize]{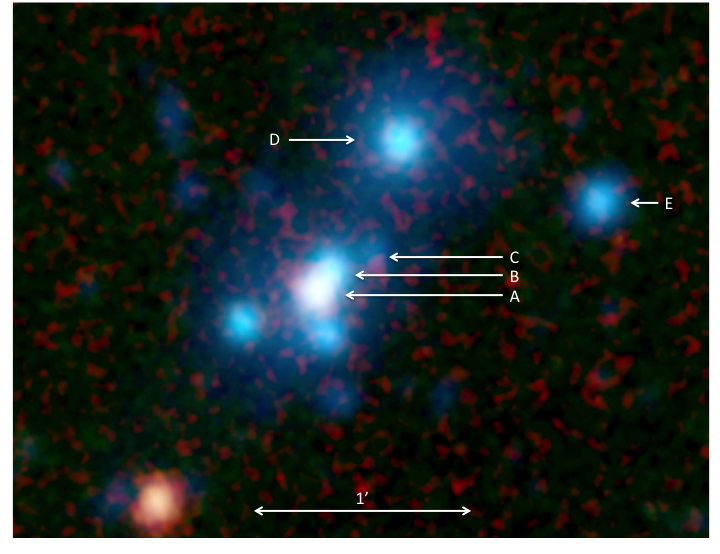}
\caption{Left: The SED fit (black line) to the WISE flux density measurements (blue points) of the GRG host (galaxy A), and the 2MASS extended source catalogue flux density measurements (green points). The WISE bands coloured blue (W1) and green (W2) depict the short wavelength bands which trace the evolved populations, and the bands coloured orange (W3) and red (W4) depict the long-wavelength bands which are sensitive to ISM heating of molecules and dust. In this way, early-type galaxies appear blue due to the dominant starlight component, while late-type galaxies and other actively star-forming galaxies will exhibit red colours. 
Right: WISE four-band view of the host system, with the colour assigned to each band as shown in the left of this figure. The GRG host galaxy is at the centre (A), with galaxies A and B moderately blended. While galaxies B--E have early-type colours, the host galaxy A has S0-type warm dust features.}
\label{figure:wise}
\end{figure*}

\subsection{Mid-Infrared Derived Mass and Star Formation}\label{sec:wise}

The host galaxy was observed in the Wide-field Infrared Survey Explorer \citep[WISE;][]{Wright2010} at mid-infrared bands (W1, W2, W3, W4) of 3.4, 4.6, 12 and 22 $\mu$m respectively. New image reconstructions, using the \cite{jarrett2012} technique that preserves the native resolution, resolved the triple galaxy system. Figure \ref{figure:wise} shows the WISE 4-band image and Spectral Energy Distribution (SED) of the host galaxy (A). The GRG host galaxy (A) stands out in Figure \ref{figure:wise} (right) due to ISM heating, presumably from star formation activity, but may also have a contribution from the central AGN. Note that at the shortest wavelengths (W1 and W2) the host galaxy (A) is moderately blended with its nearest neighbour galaxy (B) because of the relatively large angular resolution (6\arcsec). In order to separate the component emission in these bands, a de-blending technique developed for WISE imaging \citep{jarrett2013} was applied out. The resulting global flux density photometries for the host galaxy are $10.4 \pm 0.12$, $5.78 \pm 0.08$, $1.99 \pm 0.10$ and $3.82 \pm 0.51$ mJy for the W1, W2, W3 and W4 bands respectively. The formal uncertainties are 5 to 10 per cent, but due to the blending uncertainty they are likely much greater for W1 and W2. On the other hand, W3 and W4 have clean measurements because only galaxy A exhibits any dust emission.  

Nevertheless, the de-blend measurements provide a strong constraint on the stellar mass and the star formation activity for the GRG host. We use SED templates from \cite{Brown2014} to fit the WISE measurements and derive the \textit{k}-corrected flux densities (which include stellar continuum subtraction for bands W3 and W4), which are then used to compute luminosities (with luminosity distance 251.3 Mpc; Table \ref{table:u9555}). For the GRG host (A), the best-fit SED (Figure \ref{figure:wise}: left) corresponds to that of an S0-type (lenticular) galaxy, which is characterised by a massive stellar component and an infrared excess due to polycyclic aromatic hydrocarbons and warm dust emission. The host was also observed in the Two Micron All-Sky Survey \citep[2MASS;][]{2mass2006}, however we do not use these in the SED fit, and only plot them in Figure \ref{figure:wise} (left) for comparison. 

Following the prescription of equation 1 in \cite{cluver2014}, the stellar mass is derived from the W1-W2 colour of $-0.02$ mag, and the 3.4 $\mu$m mass--luminosity relation, giving a value of $\mathrm{log}(\mathrm{M}/\mathrm{M}_\odot) = 11.56 \pm 0.12 $. This mid-infrared derived mass is larger than the value estimated using optical bands (see Table \ref{table:u9555}), which may be due to the blending between galaxies A and B, or because the infrared is a better (more complete) tracer of the underlying stellar mass. In any case, the galaxy is clearly very massive. Using the longer wavelength bands, we estimate the star formation activity (using equation 4 in \citealt{cluver2014}), and derive a star formation rate (SFR) of $1.2 \pm 0.3 ~\mathrm{M}_\odot/\mathrm{year}$. The error is dominated by the scatter in the empirical relation. It is assumed that the AGN contribution is negligible, which may not be the case for the GRG host, so our SFR estimates should be considered an upper limit. However, the mid-infrared SED (Figure \ref{figure:wise}: left) does not indicate an appreciable power-law component from hot dust generated from black hole accretion, typical of QSO and radio-loud AGN (cf. \citealt{stern2012}). The derived SFR is similar to normal disc galaxies (such as the Milky Way), which suggests the star formation is significant relative to quiescent early-type galaxies, such as galaxies B and D in Figure \ref{figure:wise}.

\section{Discussion}\label{section:discussion}

\subsection{Morphology and classification}

We infer that this AGN is still being fuelled by an accretion disc due to its broad-line classification and near-constant flux density from 1.4 to 8.4~GHz. In systems where an AGN has exhausted its fuel supply one would expect the emission from the broad-line region to diminish on the timescale of years, and the radio emission from a sub-kpc jet to diminish on scales of hundreds of years (considering the light travel times). This indicates that the radio lobes are not a remnant feature of a previous period of activity, but are a result of current activity from the AGN. Considering the light travel time over the length of one of the lobes, this current phase of activity is expected to have been consistently active for at least 4 million years.

Models from \cite{kaiser_etal_1997} and observations in \cite{ishwar_2002} point towards GRGs having a lower luminosity than smaller radio galaxies, where the lobes lose energy due to both inverse-Compton and synchrotron processes. To place this GRG in the context of Figure 6 in \cite{ishwar_2002}, we assume a spectral index of -0.7, and use ${\textrm{H}}_0 = 50.0\,\textrm{km} \,\textrm{s}^{-1} \textrm{Mpc}^{-1}$. This makes its projected linear size 3.48~Mpc, and the luminosity at 1.4~GHz to be $4.33 \times 10^{24}$ W/Hz. This makes it the second largest in their sample, and indicative of a FR-II type based on its luminosity, although it is likely that selection effects in their sample limit the identification of FR-I types above 1~Mpc.

Using the luminosity cut off of $2 \times 10^{25}$ W/Hz between types FR-I and FR-II found in \cite{kaiser_etal_1997}, this GRG is classed as type FR-I. This assumes a spectral index of -0.7 to convert to 178~MHz, and a conversion to their cosmology with ${\textrm{H}}_0 = 50.0\,\textrm{km} \,\textrm{s}^{-1} \textrm{Mpc}^{-1}$, which gives a luminosity of $1.32 \times 10^{25}$ W/Hz. However, note that this is a borderline case, where there are both FR-I and FR-II lower than this.

To compare the luminosity of this GRG to the sample in Figure 1 of \cite{owen1994}, we assume a spectral index of -0.7, and convert to their cosmology (${\textrm{H}}_0 = 50.0\,\textrm{km} \,\textrm{s}^{-1} \textrm{Mpc}^{-1}$), to give a luminosity at 1.4~GHz of $4.33 \times 10^{24}$ W/Hz. This places it in the borderline region between types FR-I and FR-II. However, considering that we expect this sources giant nature to further decrease the luminosity, this suggests a classification of type FR-II.

We cannot clearly classify this GRG as an FR-I or FR-II source based on its morphology in the MSSS and archival radio data. There are no conclusive enhancements of emission from the resolution of the MSSS data (without the contribution from unassociated point sources) from which to use the standard Fanaroff-Riley classification. Furthermore, the giant nature of this source contributes to a decreased luminosity, which can skew the standard Fanaroff-Riley classification. The lack of a jet on the south-west side in NVSS does suggest that the jet may be highly relativistic, which is characteristic of FR-II sources.

\subsection{True physical size}\label{size}

The true physical size of the GRG will depend on its orientation which we assume, given its linear appearance and the low-density environment, to be aligned with its jets. The detection of the broad-line emission, in the context of unification models, implies that the inclination angle between the jets of the GRG and the line of sight is in the range 0--70$^\circ$, with a mean expected value of 33$^\circ$ \citep[][figure 2, top]{orientation2016}. The asymmetric jet brightness, with jet activity (assumed to be relativistic) detected only on the north-east side in both the NVSS and FIRST data, also suggests an inclined orientation, with the south-west jet undetected because it is Doppler-boosted away from us. We conclude that the south-west lobe is pointing away from us, and the north-east lobe towards us.

Given this orientation, and its 2.56 Mpc projected size, it is one of the largest known GRG. The largest previously published GRG, J1420-0545, has a projected size of 4.91~Mpc (\citealt{largestGRG2008} converted to our cosmology), but is believed to be aligned closely with the plane of the sky, so its true physical size is likely to be close to this value. If the GRG reported here were inclined at an angle less than 31$^\circ$ to the line of sight (close to the mean expected value), it would be larger than J1420-0545. The minimum plausible size for this GRG, given that broad-line emission is unlikely to be observed at an inclination angle above 70$^\circ$, is 2.72~Mpc.

\subsection{Low-density environment}

Whilst the galaxy distribution is not necessarily representative of the gas distribution, it is an adequate first-order approximation. The MSPM analysis of the galaxy distribution indicates a low-density environment surrounding UGC 9555, and the X-ray temperature derived from the group's velocity dispersion is consistent with this. This low-density environment allows the lobes of this GRG to grow and expand without being significantly slowed or deflected, resulting in its large physical extent. This is in line with the general trend of GRG growing in low-density environments.

The distribution of spectroscopically identified galaxies in the group is predominantly around the north-east lobe of this GRG, which is brighter. There are no spectroscopically identified galaxies in this group around the south-west lobe of the GRG. There may be a connection between this galaxy distribution and the relative brightness of the lobes of the GRG, with the north-east lobe brighter due to interaction with gas in the group, and the south-west lobe dimmer because of a lack of gas. Dedicated X-ray observations to probe the groups gas distribution would be required to confirm the interaction of the lobes with the surrounding IGM.

\subsection{Merger evidence}

The host galaxy does not show any resolved structure in SDSS, appearing as an elliptical galaxy. However, the SED fit in Figure \ref{figure:wise} indicates that the host is a lenticular galaxy, with moderate star formation activity, presumably as a result of one or more recent moderate mergers.

The SDSS data in Figure \ref{figure:sdss_smoothed_zoom} show extended optical emission around the system, which becomes prominent once each SDSS band is smoothed. We suggest that recent merger activity and tidal interactions between the nearby galaxies are causing this. Galaxy D could be the main contributor to the tidal interactions, given its orientation with the extended stellar emission, and its large size. The velocity dispersion of this group of galaxies (Table \ref{table:MSPM}; MSPM 0251) is relatively high given the low galaxy count, indicating the system is not relaxed or virialised, agreeing with other evidence that this system is undergoing merging activity.

Overall, this system agrees with evidence that galaxy mergers are the triggering mechanism for broad line radio galaxies and their radio jets \citep{review}. In this scenario, we expect the accretion mechanism to be a sudden supply of cold gas as discussed by \cite{Hardcastle2007}. In contrast to the situation in weak line radio galaxies, which are fuelled by the gradual accretion of hot gas from inside a large galaxy cluster.

Observations of the gas distribution and kinematics associated with the host galaxy would be of great interest to identify the source of AGN feeding and provide further insight to the interactions taking place in this group of galaxies. Unfortunately, the redshift of this source places the neutral hydrogen \ion{H}{i}) line at a frequency badly affected by terrestrial radio frequency interference (airport radar). Both the Effelsberg \ion{H}{i} survey \citep{ebhis2011} and the Arecibo $\alpha.40$ \ion{H}{i} source catalog \citep{a402011} were inspected, but neither a detection nor a useful limit could be provided. We suggest that forthcoming radio telescopes at remote, extremely radio-quiet sites \citep[ASKAP, MeerKAT;][]{askap2012, meerkat2011} would be very useful to investigate the gaseous distribution.

\section{Conclusions}\label{section:conclusion}
Our main conclusions in this paper are as follows:

\begin{itemize}

\item[$\bullet$] MSSS data at 142 MHz have enabled the discovery of this GRG, with a projected linear size of $2.56 \pm 0.07$ Mpc. We present evidence that this GRG is orientated out of the plane of the sky, making its true physical extent substantially larger than this.

\item[$\bullet$] We present a preliminary estimate of its integrated flux density at 142 MHz of $1.54 \pm 0.2 ~\rm{Jy}$ over the whole dual-lobe emission, including underlying background point sources, giving a total luminosity at 142 MHz of $1.16 \times 10^{25}$ W/Hz.

\item[$\bullet$] The host AGN is in a broad line galaxy, with an SED corresponding to an S0-type lenticular galaxy and with a dust-obscured SFR ($1.2 \pm 0.3 ~\mathrm{M}_\odot /\mathrm{year}$) that is significant relative to quiescent early-type galaxies.

\item[$\bullet$] The host system is a low-density galaxy group, and smoothed SDSS maps reveal the host galaxy is surrounded by extended stellar emission, indicating tidal effects and recent merger events have disturbed the system. 

\item[$\bullet$] Overall, the evidence is consistent with recent moderate mergers supplying gas to an already formed SMBH, triggering and fuelling the radio jets.

\item[$\bullet$] The MSSS and archival radio data are unable to conclusively categorise this as an FR-I or FR-II radio galaxy. Its low luminosity suggests it is a boarder-line case between FR-I and FR-II, although one needs to consider that the large size (and therefore large age) contributes towards a decreased luminosity. Further data gathered from a deep LOFAR observation are expected to clearly classify the properties of this GRG, and will be presented in a future paper (Clarke et al. in preparation).

\end{itemize}

LOFAR provides an excellent potential for the discovery of new GRG with its excellent sensitivity to large scales, as well as operating at lower frequencies where GRG will be brighter. Once the MSSS data has been fully explored we expect more discoveries of GRG. Furthermore, the LOFAR Surveys Key Science Project \citep{Rottgering2006}, is conducting a survey \citep[The LOFAR Two-meter Sky Survey;][]{shimwell2016} that is expected to produce deep low frequency observations of the entire northern sky, significantly enhancing the potential for discovery. In particularly we expect more discoveries of FR-I type, since these exhibit lower luminosities which are harder to find without utilising low frequency observations. 

\begin{acknowledgements}
AOC, JDB, TMC, DDM and AMMS gratefully acknowledge support from the European Research Council under grant ERC-2012-StG-307215 LODESTONE. MJH acknowledges support from the UK Science and Technology Facilities Council [ST/M001008/1]. BNW acknowledges support from the NCN OPUS UMO-2012/07/B/ST9/04404 funding grant. WJ acknowledges support from the Polish National Science Centre grant No. 2013/09/N/ST9/02511. We thank Martha Haynes for checking the ALFALFA data for this source. LOFAR, the Low Frequency Array designed and constructed by ASTRON, has facilities in several countries, that are owned by various parties (each with their own funding sources), and that are collectively operated by the International LOFAR Telescope (ILT) foundation under a joint scientific policy. This research has made use of the NASA/IPAC Extragalactic Database (NED) which is operated by the Jet Propulsion Laboratory, California Institute of Technology, under contract with the National Aeronautics and Space Administration. Funding for the Sloan Digital Sky Survey IV has been provided by
the Alfred P. Sloan Foundation, the U.S. Department of Energy Office of
Science, and the Participating Institutions. SDSS-IV acknowledges
support and resources from the Center for High-Performance Computing at
the University of Utah. The SDSS web site is www.sdss.org. SDSS-IV is managed by the Astrophysical Research Consortium for the
Participating Institutions of the SDSS Collaboration including the
Brazilian Participation Group, the Carnegie Institution for Science,
Carnegie Mellon University, the Chilean Participation Group, the French Participation Group, Harvard-Smithsonian Center for Astrophysics,
Instituto de Astrof\'isica de Canarias, The Johns Hopkins University,
Kavli Institute for the Physics and Mathematics of the Universe (IPMU) /
University of Tokyo, Lawrence Berkeley National Laboratory,
Leibniz Institut f\"ur Astrophysik Potsdam (AIP),
Max-Planck-Institut f\"ur Astronomie (MPIA Heidelberg),
Max-Planck-Institut f\"ur Astrophysik (MPA Garching),
Max-Planck-Institut f\"ur Extraterrestrische Physik (MPE),
National Astronomical Observatory of China, New Mexico State University,
New York University, University of Notre Dame,
Observat\'ario Nacional / MCTI, The Ohio State University,
Pennsylvania State University, Shanghai Astronomical Observatory,
United Kingdom Participation Group,
Universidad Nacional Aut\'onoma de M\'exico, University of Arizona,
University of Colorado Boulder, University of Oxford, University of Portsmouth,
University of Utah, University of Virginia, University of Washington, University of Wisconsin,
Vanderbilt University, and Yale University.
\end{acknowledgements}

\bibliographystyle{aa}
\bibliography{Clarke_MSSS_GRG}

\end{document}